\documentclass[nopreprintline,final]{elsarticle}

\usepackage{hyperref}
\usepackage{multirow}
\usepackage{amssymb} 
\usepackage{color}
\usepackage{amsmath,graphicx}

\def\c#1=#2.{\chardef#1='#2}

\c\bp=001.
\c\bpv=002.
\c\lp=003.
\c\lpv=004.
\c\bs=005.
\c\bsv=006.
\c\ls=007.
\c\lsv=010.
\c\bn=011.
\c\ln=012.
\c\bnv=013.
\c\lnv=014.
\c\pp=015.
\c\ppv=016.
\c\tp=017.
\c\tpv=020.
\c\ps=021.
\c\psv=022.
\c\ts=023.
\c\tsv=024.
\c\pn=025.
\c\tn=026.
\c\pnv=027.
\c\tnv=030.
\c\bm=031.
\c\bmv=032.
\c\lm=033.
\c\lmv=034.
\c\pm=035.
\c\pmv=036.
\c\tm=037.
\c\tmv=040.
\c\bd=041.
\c\bdv=042.
\c\ld=043.
\c\ldv=044.
\c\pd=045.
\c\pdv=046.
\c\td=047.
\c\tdv=050.
\c\bb=051.
\c\bbv=052.
\c\lb=053.
\c\lbv=054.
\c\pb=055.
\c\pbv=056.
\c\tb=057.
\c\tbv=060.
\c\fp=061.
\c\fP=062.
\c\fPv=063.
\c\fs=064.
\c\vhfw=065.
\c\vhbw=066.
\c\vhmw=067.
\c\vhfwr=070.
\c\vhbwr=071.
\c\vhmwr=072.
\c\vlfw=073.
\c\vlbw=074.
\c\vlmw=075.
\c\vlfwr=076.
\c\vlbwr=077.
\c\vlmwr=100.
\c\vmfw=101.
\c\vmbw=102.
\c\vmmw=103.
\c\vmfwr=104.
\c\vmbwr=105.
\c\vmmwr=106.
\c\vhfp=107.
\c\vhbp=110.
\c\vhmp=111.
\c\vhfpr=112.
\c\vhbpr=113.
\c\vhmpr=114.
\c\vlfp=115.
\c\vlbp=116.
\c\vlmp=117.
\c\vlfpr=120.
\c\vlbpr=121.
\c\vlmpr=122.
\c\vmfp=123.
\c\vmbp=124.
\c\vmmp=125.
\c\vmfpr=126.
\c\vmbpr=127.
\c\vmmpr=130.
\c\gbreath=157.			
\c\gvoice=131.
\c\ground=132.
\c\gthroat=133.
\c\gback=134.
\c\gbackround=135.
\c\glip=136.
\c\glipback=137.
\c\gpoint=140.
\c\gpointround=141.
\c\gtop=142.
\c\gtopround=143.
\c\minner=144.
\c\mouter=145.
\c\mclose=146.
\c\mopen=147.
\c\mtrill=150.
\c\mnasal=151.
\c\moneside=152.
\c\mbothsides=153.
\c\minverted=154.
\c\mprotruded=155.
\c\mstop=156.
\c\memission=157.
\c\memissionstopped=160.
\c\msuction=161.
\c\msuctionstopped=162.
\c\mhiatus=163.
\c\mabrupt=164.
\c\mholder=165.
\c\maccent=166.
\c\memphasis=167.
\c\mwhistle=170.
\c\mvwhistle=171.
\c\mplus=172.
\font\vis=vispeech 

\def\mycopyright{\copyright 2021. This manuscript version is made available under the
CC-BY-NC-ND 4.0 license http://creativecommons.org/licenses/by-nc-nd/4.0/. Full
article at DOI: 10.1016/j.csl.2021.101213}
\newtheorem {theorem} {Experiment}

\def\mymixer{\alpha}
\def\mydot{\circ}
\def\myhat{\wedge}
\def\Healthy{{\tt healthy}}
\def\Dysarthric{{\tt dysarthric}}
\def\Low{{\tt low}}
\def\High{{\tt high}}
\def\Medium{{\tt medium}}
\def\unk{$<${\tt unk}$>$}
\def\nmunk{<{\tt unk}>}

\def\SS{W}
\def\DS{{\tt{DeepSpeech}}}
\def\VeryLow{{\tt{very low}}}
\def\F{{\vec{F}}}
\def\d{\delta}
\def\h{\eta}
\def\myspace{\square}
\def\myquote{\star}
\def\N{|W|}
\def\T{{\cal T}}
\def\r{|W_k|}
\def\UW{{\tt Uncommon Words}}
\def\CW{{\tt Common Words}}
\def\Digits{{\tt Digits}}
\def\Letters{{\tt Letters}}
\def\CC{{\tt Computer Commands}}
\def\C{{\cal C}}
\def\D{{\cal D}}
\def\S{{\cal S}}
\def\L{{\cal L}}
\def\PC{{{\sc Pc}}}

\def\NATURALIZATION{{\tt Naturalization}} 
\def\AUTOBIOGRAPHY{{\tt Autobiography}}
\def\EXACTITUDE{{\tt Exactitude}} 
\def\IRRESOLUTE{{\tt Irresolute}} 
\def\INALIENABLE{{\tt Inalienable}} 
\def\LEGISLATURE{{\tt Legislature}}
\def\OVERSHADOWED{{\tt Overshadowed}}
\def\PSYCHOLOGICAL{{\tt Psychological}} 
\def\DISSATISFACTION{{\tt Dissatisfaction}}
\def\AGRICULTURAL{{\tt Agricultural}}
\def\APOTHECARY{{\tt Apothecary}}
\def\AUTHORITATIVE{{\tt Authoritative}}
\def\EXAGGERATE{{\tt Exaggerate}}
\def\INEXHAUSTIBLE{{\tt Inexhaustible}}
\def\naturalization{\NATURALIZATION}
\def\autobiography{\AUTOBIOGRAPHY}
\def\exactitude{\EXACTITUDE}

\def\overshadowed{\OVERSHADOWED}
\def\psychological{\PSYCHOLOGICAL} 
\def\dissatisfaction{\DISSATISFACTION} 
\def\agricultural{\AGRICULTURAL} 
\def\apothecary{\APOTHECARY}
\def\authoritative{\AUTHORITATIVE} 

\def\inexhaustible{\INEXHAUSTIBLE}

\def\mycomment#1{{}}
\def\alphabetset{{\cal A}}
\def\mydef{\stackrel{\Delta}{=}}
\def\mychange#1{{{#1}}}
\def\E{{\mychange{E_{word}}}}

\journal{Computer Speech and Language}

\begin{document}

\begin{frontmatter}

\title{Automatic Speaker Independent Dysarthric Speech Intelligibility Assessment System\footnote{\mycopyright}}

\author{Ayush Tripathi, Swapnil Bhosale, Sunil Kumar Kopparapu}
\address{TCS Research and Innovation - Mumbai, Tata Consultancy Services
Limited, India.}

\ead[url]{www.tcs.com}

\ead{\{t.ayush,bhosale.swapnil2,sunilkumar.kopparapu\}@tcs.com}

\begin{abstract}
Dysarthria is a condition which hampers the ability of an individual to 
control the muscles that play a major role in speech delivery. The loss 
of fine control over muscles that assist the movement of lips, vocal chords, 
tongue and diaphragm results in abnormal speech delivery. One can assess the
severity level of dysarthria by analyzing the intelligibility of speech spoken
by an individual. Continuous intelligibility assessment helps speech 
language pathologists not only study the impact of medication but also 
allows them to plan personalized therapy. It helps the clinicians immensely 
if the intelligibility assessment system is reliable, automatic,  simple for
(a) patients to undergo and (b) clinicians to interpret. Lack of availability
of dysarthric data has resulted in development of speaker {\it dependent} 
automatic intelligibility assessment systems which requires patients to speak a
{\it large} number of utterances. In this paper, 
we propose (a) a cost minimization procedure to select an optimal 
({\it small}) number of utterances that need to be spoken by the dysarthric patient,
(b) four different speaker {\it independent} intelligibility 
assessment systems which require the patient to speak a small number of words,
and (c) the assessment score
is close to the perceptual score that the \mychange{Speech Language Pathologist (SLP)}
can relate to.
The need for small number of utterances to be spoken by the patient and the
score being relatable to the SLP benefits both the dysarthric patient and the
clinician from usability perspective.
\end{abstract}


\end{frontmatter}



\section{Introduction}
\label{sec:introduction}

Dysarthria is a condition which hampers the ability of a person to control the 
muscles that play a major role in speech delivery. The loss of fine control 
over muscles that assist the movement of lips, vocal chords, tongue and diaphragm 
results in abnormal speech delivery. Among other things, this  condition
affects verbal communication because of hampered intelligibility of spoken
speech.

It is believed that $7$ of the $12$ cranial nerves, 
which perform the task of sending sensory and motor information to all the 
muscles of the body, are 
related to {\it speech} and language. Damage to any of
these nerves can result in hampering control of muscles that assist in 
speech production \cite{book:894458}.
Generally dysarthria is a result of a disease or a
stroke that affects the nervous system and not a disease in itself. 
The speech production is a complex mix of respiration,
phonation, resonance, articulation, and prosody. If any of these speech process
is affected because of lack of muscle control speech production suffers.
Depending on the degree of loss of control
over the muscles that assist in articulation dysarthria can range from mild
where
a person sounds as good as a healthy speaker to severe where it might be very
difficult for a listener to understand what is being spoken. There are $6$
types of dysarthria, namely, (a) Ataxic, (b) Flaccid, (c) Hyperkinetic, (d) 
Hypokinetic, (e) Spastic and (f) Mixed (a mix of Flaccid and Spastic dysarthria). 
\begin{table}
\caption{Causes for Different Dysarthria \cite{book:894458}.}
\label{tab:dysarthria_causes}
\begin{center}
\scalebox{0.5}{
\begin{tabular}{|l|c|c|c|c|c|c|c|}\hline
& \multicolumn{6}{c|}{Dysarthria Type}\\ \cline{2-7}
Possible Causes&Ataxic&Flaccid&Hyperkinetic&Hypokinetic&Mixed 
&Spastic\\ \hline
Stroke (Cerebro Vascular Accident)&\checkmark&\checkmark&\checkmark&&\checkmark&\checkmark \\
Trauma&\checkmark&\checkmark&&&\checkmark&\checkmark \\
Tumor&\checkmark&\checkmark&\checkmark&&&\checkmark \\
Congenital conditions&\checkmark&\checkmark&&&&\checkmark \\
Infection&\checkmark&&&&&\checkmark \\
Amyotrophic lateral sclerosis&&&&&\checkmark& \\
Athetosis&&&\checkmark&&& \\
Ballism&&&\checkmark&&& \\
Chorea Infection&&&\checkmark&&& \\
Drug-induced Dyskinesia&&&\checkmark&&& \\
Dystonia&&&\checkmark&&& \\
Gilles de la Tourette's syndrome&&&\checkmark&&& \\
Palsies&&\checkmark&&&& \\
Parkinsonism Drug-induced&&&&\checkmark&& \\
Toxic effects&\checkmark&&&&& \\
Viral Infection&&\checkmark&&&& \\ \hline
\end{tabular}
}
\end{center}
\end{table}

The cause for different types of dysarthria is well studied in literature \cite{book:894458}.
As seen in  Table \ref{tab:dysarthria_causes}, cerebrovascular accident (CVA), commonly also known as stroke, 
can result in any kind of dysarthria except, Hypokinetic dysarthria,
whereas drug induced Parkinsonism can only cause Hypokinetic dysarthria. Trauma
and Tumor can also result in a variety of dysarthria types (see Table
\ref{tab:dysarthria_causes}). While Toxic effects
can only induce Ataxia, Viral Infection causes Flaccid dysarthria, diseases like
Athetosis, Ballism, Chorea Infection, Drug-induced Dyskinesia, Dystonia and
Gilles de la Tourette's syndrome are known to result in Hyperkinetic dysarthria.

\begin{table}
\caption{Different Dysarthria effecting Speech Characteristics \cite{book:894458}.}
\label{tab:dysarthria_vs_speech_characteristics}
\begin{center}
\scalebox{0.5}{
\begin{tabular}{|l|c|c|c|c|c|c|c|}\hline 
& \multicolumn{6}{c|}{Dysarthria Type}\\ \cline{2-7}
Speech Characteristics&Ataxic&Flaccid&Hyperkinetic&Hypokinetic&Mixed 
&Spastic\\ \hline
Imprecise consonants&\checkmark&\checkmark&\checkmark&\checkmark&\checkmark&\checkmark \\
Monopitch&&\checkmark&\checkmark&\checkmark&\checkmark&\checkmark \\
Monocloudness&&&\checkmark&\checkmark&\checkmark&\checkmark \\
Distorted vowels&\checkmark&&\checkmark&&\checkmark& \\ 
Harsh voice (quality)&\checkmark&&\checkmark&\checkmark&\checkmark&\checkmark \\
Hypernasality&&\checkmark&&&\checkmark&\checkmark \\
Excess and equal stress&\checkmark&&&&\checkmark& \\
Irregular articulatory breakdowns&\checkmark&&\checkmark&&& \\
Low pitch&&&&&\checkmark&\checkmark \\
Reduced stress&&&&\checkmark&&\checkmark \\
Short phrases&&&&&\checkmark&\checkmark \\
Slow rate&&&&&\checkmark&\checkmark \\
Strained-strangled voice&&&\checkmark&&&\checkmark \\ 
Breathy voice&&\checkmark&&\checkmark&& \\ 
Inappropriate silences&&&&\checkmark&& \\
Loudness control problems&\checkmark&&&&& \\
Nasal emission&&\checkmark&&&& \\
Prolonged intervals&&&&&\checkmark& \\
Short rushes of speech&&&&\checkmark&& \\
Variable nasality&\checkmark&&&&& \\ \hline
\end{tabular}
}
\end{center}
\end{table}
\mychange{A comprehensive  mapping between different types of dysarthria and the speech characteristics in shown in Table \ref{tab:dysarthria_vs_speech_characteristics}.}
\mychange{Out of the 20 speech characteristics, the inability to precisely articulate consonants is a prime characteristic across all types of dysarthria.}
One of the prime speech characteristic, across all type of dysarthria is the 
inability to articulate consonants precisely (see Table
\ref{tab:dysarthria_vs_speech_characteristics}). As can also be seen multiple
speech characteristics are visible in a particular type of dysarthria. Clearly a
combination of these speech characteristics make the speech of a dysarthric
patient unintelligible. The speech characteristics become increasingly visible,
rather audible, as dysarthria becomes more profound in a patient. For example, 
the
degree of distortion of a vowel ("vowel distortion") becomes more profound as dysarthria progresses
from mild to severe in case of Ataxic and Hyperkinetic dysarthria. In case of 
Mixed dysarthria, there is an increase in "Prolonged interval" between the words or
phonemes in spontaneous speech when severity level of dysarthria increases
\mychange{thus, further degrading the intelligibility.}

Some of the noticeable characteristics of dysarthria are 
(a) there is never an island of clear speech,
the speech errors are seen uniformly along the entire speech utterance,
\mychange{(b) articulation error is caused due to distortions and deletions, and not the insertion of phonemes,}
(c) pronunciation
of consonants are consistently imprecise, (d) vowels are neutralized,
(e) the speech delivery rate is slow and labored 
and 
(f) 
any word requiring large articulatory movement due to complexity in 
pronunciation results in a decreased articulatory 
performance.

A speech language
therapist (SLP) who specializes in speech therapy can help a person with
dysarthria improve speech delivery through medication and practice of suitable exercises to regain control over the articulators.
\mychange{However, this requires }an accurate assessment of the degree of dysarthria 
at the time of diagnosis and during therapy\mychange{,} to understand the effect of
medication.

Instrumental investigation
\mychange{for example, using} 
a water manometer fitted with a bleed valve 
(to measure sub-glottal air pressure),
laryngograph 
(measure abnormalities of closure),
electropalatography 
(tongue movement), pneumotachography (to measure differentials of
nasal and oral air flow) \cite{book:1067226} 
are supplemented 
by perceptual (human) assessment and in many cases, due to lack of instruments,
perceptual assessment might be the only possible means of evaluation possible. 
Perceptual assessment by a trained 
SLP is considered the gold standard even if we were not to consider the
fact that 
instrumentation based approach are invasive, expensive and  painful.

Several tests and metrics developed over the years are used by the
SLPs.  Three frequently used tests for evaluating dysarthric speech are 
(a) Assessment of Intelligibility of Dysarthric Speech \cite{AIDS}, 
(b) Frenchay Dysarthria Assessment \cite{FDA}, and (c)
Quick Assessment for Dysarthria \cite{bk_qad}. While
Hoehn and Yahr scale \cite{HnY}, 
Unified Parkinson's Disease Rating Scale (UPDRS) \cite{UPDRS},
and Scale for Assessment and Rating of Ataxia (SARA) \cite{SARA} are 
frequently used to measure the severity of dysarthria. 
All these metrics are influenced 
by the type of stimuli used to elicit a phonetic utterance 
\cite{Kent}. For example, the assessment of speech part of the UPDRS has a
scale between $0$ and $4$, where $0$ is normal and a score of $4$ suggests 
that the speaker is unintelligible most of the time. 
As seen in Table \ref{tab:updrs} the interpretation of 
{\em mildly}, {\em moderately}, {\em severely} 
are not only SLP ({\em evaluator}) dependent but, more importantly, these 
interpretations have a bearing on the stimuli used to elicit speech from the
patient.

\begin{table}

\caption{UPDRS for Speech Intelligibility Assessment \cite{updrs_metric}. The underlined words 
 are subject to human interpretation.}
\begin{center}
\scalebox{0.8}{
\begin{tabular}{|c|l|} \hline
UPDRS & Interpretation \\ \hline
0 & Normal \\ 
1 & \underline{Mildly} affected.  No difficulty being understood. \\
2 & \underline{Moderately} affected. \underline{Sometimes} asked to repeat statements. \\
3 & \underline{Severely} affected.  \underline{Frequently} asked to repeat statements. \\ 
4 & Unintelligible \underline{most} of the time.\\ \hline
\end{tabular}
}
\end{center}
\label{tab:updrs}
\end{table}

Both objective assessment using instruments and perceptual assessment by
clinicians show drawbacks. As a result, the research focus is on using 
signal processing and machine learning \mychange{(ML)} approaches to automatically assess 
dysarthric speech intelligibility. 
In this paper we concentrate on automatic dysarthria speech
intelligibility assessment. This paper consolidates our work reported in \cite{9053339}, \cite{9054492} 
and \cite{icds2020} and expands on it in the following way, (a) 
\mychange{we}
 propose two
additional new methods to robustly compute the speech intelligibility of the speaker, (b) we derive using visible speech the characteristics of words that make them useful to be used for speech intelligibility characterization, (c) we propose a method to enable selection of
an optimal number of words that are sufficient for intelligibility assessment for a language from a set of dictionary
words (vocabulary) without affecting the correctness of the assessed intelligibility score.  
The rest of the paper is organized as follows, in Section
\ref{section:review} we review the existing literature on intelligibility
assessment techniques elaborating on the perceptual, instrumental and automatic intelligibility assessment techniques. 
We describe the proposed intelligibility assessment techniques in Section \ref{sec:proposed} and also
describe the method to derive an optimal set of words from a set of dictionary words in Section \ref{sec:word_selection}. 
The experimental setup is described in Section \ref{sec:experiments} and we conclude in Section \ref{sec:conclusions}.

\section{Review of Intelligibility Assessment Methods} 
\label{section:review}

The approaches adopted in literature for speech intelligibility assessment can be 
classified into three broad categories, namely, 
(a) physiological assessment using sophisticated instruments, 
(b) perceptual assessment performed by a trained clinician, 
and
(c) automatic speech intelligibility assessment using advancement in 
signal processing and \mychange{ML.}

The current research focus as well as the thrust of this paper 
is on the use of automatic methods for dysarthric
speech intelligibility assessment. We carry out a brief survey of the existing 
techniques and not review the perceptual or instrument
based assessment techniques.

The advancement in signal processing and \mychange{ML}
literature has encouraged researchers to explore the use of signal processing and ML techniques for automatic speech intelligibility assessment.
Broadly, these methods aim at measuring the abnormalities in spoken 
speech by extracting handcrafted acoustic features based on statistical signal 
processing \cite{Acoustic,spmethods2} and/or supervised methods based on 
ML \cite{ML,7953122}. Such techniques offer the advantage of frequent, 
cost effective and objective assessment of speech intelligibility. 
However, a speech signal is abundant in layered information such as 
gender, speaker traits, emotion of the speaker in addition to the linguistic
content of the speech \cite{book:1291324}. These multiple layers (who spoke, how
they spoke, what they spoke) of information in the same speech sample makes 
it difficult to extract features that carry {\em only} pathology-specific 
information. 
Additionally, the ML models are prone to overfitting during training because
of the small corpus size of the disease-specific speech. This makes these
models ineffective in terms of generalization over a larger 
population. One set of such approaches aims at categorizing the patients speech 
recording into broad categories such as $\{\Healthy, \Dysarthric\}$ ($2$
classes)
\cite{kim2015automatic,rudzicz2010articulatory} or $\{\High, \Medium, \Low\}$ Intelligibility ($3$ classes) rather than 
providing an absolute intelligibility assessment score.  
However, a mechanism that can provide continuous intelligibility ratings is of 
more use in a clinical setting especially, when the assessment scores are \mychange{synchronous}
with the perceptual scores understood by the SLP. 
Subsequently, a set of approaches aimed at learning continuous intelligibility
assessment scores have been been researched. 
Literature in this area can be broadly categorized into (a) reference-free and
(b) reference-based approaches.

\subsection{Reference-free approaches}

Reference-free approaches aim at measuring speech intelligibility without using 
any prior knowledge associated with healthy or intelligible speech as reference. 
Different handcrafted features that are believed to be correlated 
with speech intelligibility have been explored  to 
(a) classify different types of dysarthria 
(Table \ref{tab:dysarthria_vs_speech_characteristics})
(b) assess intelligibility of speech especially that affected by dysarthria.

Phonation features that describe pathological voice such as fundamental 
frequency $F_0$ and jitter have been found useful
in quantification of voice tremor \cite{jitter,tremor}.
Pitch Period Entropy based assessment was proposed \cite{ppe} in order to 
overcome the gender and acoustic environment dependency of these features. 
Short-time energy and variation of energy (shimmer) \cite{shimmer} have also 
been effectively used to describe hypophonia. Teager-Kaiser Energy Operator 
\cite{tkeo}, a measure of instantaneous speech intensity, has been used in 
order to take signal frequency into account. Features that capture energy 
distribution in power spectra such as Median of Power Spectral Density (MPSD) 
\cite{mpsd} and Low Short-Time Energy Ratio (LSTER) \cite{lster} have also 
been explored in literature. 
Acoustic cues based on 
first three formants, 
and their corresponding bandwidths 
can be observed to study the impact on articulatory dynamics, thereby proving to be 
helpful in estimation of speech intelligibility \cite{formant}. Vowel Space 
Area (VSA) 
for studying speech 
intelligibility \cite{vsa1, vsa2} has also been
explored. 

An approach 
for discriminating dysarthric speech from healthy speech by using a set of 
glottal 
and openSMILE features have been explored using  
Support Vector Machine classifier \cite{Narendra}. 
An investigation of analytic phase features, 
extracted from the speech signal, by using single frequency filtering technique was performed in \cite{Krishna}. 
Audio descriptor features used for defining Timbre of musical instruments 
along with \mychange{Artificial Neural Network (ANN)} model was used in \cite{chitralekha} to classify severity levels
of dysarthric speech.
Multi-tapered spectral estimation 
to obtain the audio descriptor features was employed for dysarthria classification. 
A multi-task learning technique to jointly solve dysarthria detection and 
speech reconstruction tasks was explored by encoding dysarthric speech to a 
lower dimensional latent space in \cite{daniel}. 
Speech rate, pauses, fillers, and Goodness of Pronunciation (GoP) were
used as discriminating features to differentiate healthy controls (HC) from individuals with Huntington 
disease using \mychange{Long Short Term Memory - Recurrent Neural Network (LSTM-RNN)} and \mychange{Deep Neural Network (DNN)} \cite{Perez}. 
Classification of patients with \mychange{Amyotrophic Lateral Sclerosis (ALS)}, \mychange{Parkinsons Disease (PD)}, and \mychange{Healthy
Control (HC)} using a \mychange{Convolutional Neural Network - Long Short Term
Memory (CNN-LSTM)} based transfer 
learning framework was proposed in \cite{jhansi}. 
Dysarthria detection in Mandarin speaking individuals was proposed in
\cite{mayle} using a RNN-LSTM based framework directly on raw speech waveforms. 
Variations of CNN architecture such as time-CNN, frequency-CNN and tf-CNN to 
capture spectro-temporal variations in speech of individuals suffering from dysarthria was explored in \cite{chandrashekhar}. 
The performance of Bi-directional LSTM (BiLSTM) with log-filterbank, 
Mel-filter Cepstral Coefficients (MFCCs) and i-vector features as input to classify Dutch and 
English speakers into intelligible and non-intelligible categories was explored 
in \cite{chitra-jstsp}. 
CNN for automatic early detection of ALS from highly intelligible speech was
attempted in \cite{An2018}. 
The use of features based on occurrence of \unk\ token 
in \DS, an end-to-end speech-to-alphabet system based on the 
\mychange{CTC} loss function was effectively employed in 
\cite{9054492} in order to achieve a $4$-class classification of different 
intelligibility levels of dysarthria.

These reference-free methods rely on supervised learning and are usually trained 
on a small dataset. Due to the size of pathological speech corpora (low
resource), they are likely to overfit the train data, thus not performing
very well on the test (unseen during training) dataset. In order to tackle this 
situation, more recently, several reference-based approaches have been proposed.

\subsection{Reference-based approaches}

The family of reference-based or non-blind approaches involve the use of 
healthy reference signals in a wide variety of ways. The reference speech signals 
are used to train systems for example, \mychange{Automatic Speech Recognition (ASR)}
engines, 
which are then employed for evaluating pathological speech to 
estimate their intelligibility with respect to the healthy speech.

A single speaker-independent Gaussian Mixture Model (GMM) is trained on the data of 
healthy speakers to create a healthy reference model in \cite{tobias}. The 
reference GMM is adapted using pathological speech to generate a GMM-based 
supervector (feature) 
to represent the pathological speech signal. The intelligibility score 
is then obtained by training a regression model on the GMM-based supervector. 
A total variability subspace modeled by factor analysis method was adopted
to assess dysarthria intelligibility 
in
\cite{Martinez}. Acoustic information corresponding to each speech recording was 
represented by an i-vector and a support vector regression model was trained 
for intelligibility score estimation. 
A very similar approach of using i-vector for regression task has been adopted 
in both \cite{martinez2} and \cite{imed}. GMM and DNN based models trained using 
MFCC features were employed for the task of hypernasality estimation in 
\cite{vikram}. 
Two acoustic models trained on a large corpus of 
healthy speech, one to measure the nasal resonance from voiced sounds and another
to measure the articulatory 
imprecision from unvoiced sounds was used in \cite{Saxon} to estimate hypernasality in 
dysarthric subjects.  

A phonological feature extractor trained using healthy speech samples was 
employed to compute statistical phonological characteristics of a speaker 
using frame-level phonological features in \cite{middag1}. Similarly, a study 
was carried out in \cite{middag2} to understand if ML 
models trained on normal
healthy speech can be adapted to train \mychange{using} pathological speech. 

Another set of reference-based approaches are based on training an 
ASR system using healthy reference speech. 
The ASR system replaces a human listener and pathological speech intelligibility 
is computed based on the word error recognition rate. Such an approach has been 
applied to measure intelligibility of tracheoesophageal speech \cite{schuster}, 
speech of oral cancer patients \cite{maier}, head 
and neck cancer patients \cite{maier2}, and intelligibility of substitute speech after 
laryngectomy \cite{schuster2}.

More recently a short-time objective intelligibility (STOI) approach was
proposed in \cite{ICASSP2020janbakhshi}. First
utterance-dependent reference signal from multiple 
healthy speakers was constructed using Dynamic Time Warping (DTW).
 This reference and the pathological speech (corresponding to the same
utterance)  was aligned to 
compute the short-time or spectral cross-correlation. 
This method, called  P-STOI, was evaluated on French and English speakers. 
Subsequently in \cite{ICASSP2020janbakhshi}, an improvised method was proposed which 
used synthetic speech generated by a text-to-speech (TTS) systems to create 
a reference speech signal. 
Spectral bases of  the  octave  band  representations  of speech was exploited
in \cite{subspace} by first finding subspaces of spectral patterns 
characterizing intelligible (healthy) and pathological speech using 
Principal Component Analysis (PCA) or Approximate Joint Diagonalization (AJD) 
and then measuring the Grassman distance between the two subspaces.

While the automatic speech intelligibility of pathological speech has attracted
a lot of attention leading to a researchers trying out different approaches,
the main constraint has been in terms of the size of the database, which is
often very small and because of privacy issues often not available for use
by other groups.

\section{Proposed Intelligibility Assessment Approaches}
\label{sec:proposed}

We describe two approaches, one based on high level descriptors extracted
directly from speech utterances and the other based on the output of an 
end-to-end speech to alphabet (S2A) recognition engine, to assess the intelligibility of dysarthric speech. We
describe them in detail.

\subsection{High Level Descriptors based Intelligibility Assessment} \label{section:hld}

The openSMILE toolkit \cite{openSMILE} is an open-source toolkit, used for 
extraction of features from audio signals. 
OpenSMILE features have been popularly used for detection of 
emotions in audio \cite{dumpala2019a}, speaker biometric, and detection of  
voice pathologies \cite{bhat2019a}. 
We use the Interspeech 2009 Emotion Challenge configuration \cite{Schuller2009TheI2} to obtain a set of $384$ 
features corresponding to an input speech signal.
Thus, a speech signal can be represented as a $384$ dimensional vector ($F$).
\mychange{Further we normalize $F$ along each of the $384$ dimensions so that the
feature value is in the range $(0,1)$.}

Let, $\F^{k}(w)$ denote the \mychange{normalized} high level descriptors (feature vector) of 
length $|F| = 384$ corresponding to the
Speaker $k$. Note that $$E(\d,\h,w) = \mychange{\frac{1}{|F|}}\sqrt{\sum_{i=1}^{|F|} \left (\F_{i}^{\h}(w) -
\F_{i}^{\d}(w)\right )^2}$$ is the Euclidean distance between speaker $\d$ and
speaker $\h$ speaking the same word $w$ \mychange{(note that $E(\d,\h,w)$ takes a value
between $(0,1)$)}.  
Then, the intelligibility measure of speaker $d$ speaking the word $w$, denoted by $I_{os}(d,w)$ can be 
computed as 
\[
I_{os}(d,w) = \frac{1}{N_h} \sum_{l=1}^{N_h} E(d,l,w)
\]
where $N_h$ denotes the number of healthy speakers available in the dataset.
Note that $I_{os}(d,w)$ captures the average distance of the speaker $d$
speaking the word $w$ from all the $N_h$ healthy speakers who speak the same
word $w$.
The overall intelligibility of the speaker $d$ across all the words in the
dataset can be computed as
\begin{equation}
I_{os}(d) = \frac{1}{|W|} \sum_{w \in W} I_{os}(d,w)
\label{eq:i_os}
\end{equation}
where $W$ is the set of words and $|W|$ is the total
number of words in the database.
A large value of intelligibility measure $I_{os}(d)$ would mean a larger distance between the speaker
$d$ and the healthy speakers in the high level descriptors space. 
We can easily see that the metric $I_{os}$ is negatively correlated with the perceptually assessed intelligibility score
and this is clearly seen in our experimental results.

\subsection{Speech to Alphabet (\DS) based Intelligibility Assessment}

Mozilla's DeepSpeech \cite{MozillaDS} (\DS) is an end-to-end deep learning model 
that converts
speech into alphabets based on the Connectionist Temporal Classification (CTC) loss function. The $6$ layer deep model is pre-trained on $1000$ hours of speech from 
the Librispeech corpus \cite{panayotov2015librispeech}. All the layers, except
the $4^{th}$ layer, which has recurrent units, have feedforward dense units.
A speech utterance $x$ is segmented into $T$ frames, as is common in speech
processing, namely,  $x_{t}\ \ \forall t \in \left[0, T-1\right]$. 
Each frame $x_{t}$ is represented by $26$ 
\mychange{MFCCs}, namely, $\vec{f}_t$. 
Subsequently the speech utterance $x$ can be represented as a 
sequence of speech features, namely, $\{\vec{f}_t\}_{t=0}^{T-1}$. The input to
the \DS\ is the speech features from $9$ preceding and $9$ succeeding frames,
namely $\{ \vec{f}_{t-9} \cdots \vec{f}_{t+9}\}$. 
The output of the \DS\ model is a probability distribution over the
alphabet set $\alphabetset$ of a  particular language. 
In case of English language, $\alphabetset = (a, b, \cdots, z, \myspace, \myquote, \nmunk)$ and
$|\alphabetset| = 29$. Note that there are three additional
outputs, namely, \unk, $\myspace$, and $\myquote$ corresponding to {\em unknown}, {\em space} and 
an {\em apostrophe} respectively in the alphabet set in addition to the $26$ known English letters.
The output at each frame/timestep, $t$ is 

\begin{equation}
c^*_{t} = \max_{\forall  k \in \alphabetset} P\left(\left(c_{t} = k\right)|\left \{\vec{f}_{t-9}, \cdots, \vec{f}_{t}, \cdots, \vec{f}_{t+9}\right \}\right) 
\label{eq:mozds}
\end{equation}
where $c^*_{t} \in \alphabetset$. It is important to note that a typical
speech recognition engine is assisted by a statistical language model (LM)
which helps in masking small acoustic mispronunciations. However as seen in 
(\ref{eq:mozds}) there is no role of LM.
 This, as we will see
later, is vital to speech intelligibility estimation.

Let us define four string operations, namely, (a) $\C(s,p)$ which counts the
number of patterns $p \in \alphabetset$ in the string $s$, (b) $\L(s)$ is the
length of the string $s$, (c) $\S(s)$ which compresses the
string $s$ so that all repetitions are deleted and (d) $\D(s,p)$ which deletes
all the patterns $p \in \alphabetset$ in the string $s$. The string $s$ is
typically the output of (\ref{eq:mozds}). Table \ref{tab:operations} shows some
examples of these operation on the output of \DS\ string $s$. While we
will not be making use of these properties in this paper, it is interesting to
note that (a) the order of $\S$ and $\D$ does not matter and 
(b) $\S^n =\S$ and $\D^n = \D$ (see Table \ref{tab:operations}).

\begin{table}
\caption{Example operation on the output of \DS.}
\label{tab:operations}
\begin{center}
\scalebox{0.8}{
\begin{tabular}{rl} \hline
\multicolumn{2}{c}{$s = [$ {\tt n\ a\ a\ $\myspace$ t\ t\ t\ \unk\ u\ u\ u\ $\myspace$ r\ r\ r\ r\ \unk\ e\ e\ \unk} $]$}\\
\hline
$\C(s,\nmunk)=$ & $3$\\  
$\L(s)=$ & $20$\\
$\S(s)=$ & $[$ {\tt n\ a\ $\myspace$ t\ \unk\ u\ $\myspace$ r\  \unk\ e\ \unk} $]$ \\
$\D(s,\nmunk)=$ &  $[$ {\tt n\ a\ a\ $\myspace$ t\ t\ t\ u\ u\ u\ $\myspace$ r\ r\ r\ r\ e\ e} $]$ \\ \hline
$\S(\D(\D(s,\nmunk),\myspace)) =$ &$ [$ {\tt n\ a\ t\  u\  r\  e} $]$\\
$\S(\D(\D(s,\myspace),\nmunk)) = $ &$ [$ {\tt n\ a\ t\  u\  r\  e} $]$\\
$\D(\D(\S(s),\myspace),\nmunk) = $ &$ [$ {\tt n\ a\ t\  u\  r\  e} $]$\\
$\D(\S(\D(s,\myspace),\nmunk)) = $&$ [$ {\tt n\ a\ t\  u\  r\  e} $]$\\ \hline
$\S^n(s) =$ & $ \S(s)$ \\
$\D^2(s,p) \mydef \D(\D(s,p),p) =$ & $\D(s,p)$ \\ \hline
\end{tabular}
}
\end{center}
\end{table}

In order to estimate the intelligibility, for a given spoken word, $u$, we
get the output $c^*_t$ from (\ref{eq:mozds}) for each time step $t$. Let $c^* =
c^*_1, c^*_2, \cdots$
denote the output sequence for the spoken word $u$.
We compute the similarity between 
$s_1 = \D(\S(c^*),\nmunk)$ of length $l_1 = L(s_1)$ and the ground truth (say $s_{1g}$ of length $l_g
= \L(s_{1g})$). Note that the operations $\S$ and $\D$ are not applicable on the
ground truth. 
The higher the similarity between the two strings ($s_1$ and $s_{1g}$) the higher the intelligibility 
of the utterance $u$.  
We experiment with three string comparison metrics, namely, (a) Sequence Matcher,
(b) Levenshtein distance and (c) occurrence of \unk\ for computing the intelligibility scores.

\noindent {\bf Intelligibility based on Sequence Matcher:}
Sequence matcher is an extension of the Gestalt Pattern Matching algorithm \cite{black2004ratcliff}. 
The similarity between two strings is computed as the number of characters in the
matching subsequences divided by 
the total number of characters in both the strings. 
The number of matching characters is the 
length of the longest matching (contiguous) subsequence between the two input strings $s_1$ and $s_{1g}$.
The intelligibility using the Sequence Matcher technique can be computed as 
\begin{equation}
I_{sm}(s_1, s_{1g}) = 100 * \left(\frac{2*m}{l_1 + l_g} \right)
\label{eq:i_sm}
\end{equation}
where $m$ denotes the total number of characters in matching subsequences between $s_1$ and $s_{1g}$.

\noindent {\bf Intelligibility based on Levenshtein Distance:}
Levenshtein or Edit distance expresses the number of edits (insertions, deletions, 
substitutions) necessary to convert one sequence to  another. The lesser the number
of edits the higher is the similarity between the two sequences. 
The intelligibility using the Levenshtein distance is computed as,
\begin{equation}
I_{ld}(s_1, s_{1g}) = \left(1 -  \frac{l}{l_1 + l_g} \right) *100
\label{eq:i_ld}
\end{equation}
where $l$ represents the edit distance between $s_1$ and $s_{1g}$. 

\noindent {\bf Intelligibility Assessment based on \unk:} The \DS\
	speech to character has been, like most automatic speech recognition engines, 
	trained on {\em only} healthy speech and it is natural that it fails to recognize sounds
	in pathological speech, an effect of this is the production of the character
	\unk\ in (\ref{eq:mozds}). Intelligibility of the speech can be estimated from
	the number of times \unk\ occurs in the output, namely in $s_1$. 
The intelligibility using this metric is computed as,

\begin{equation}
I_{\nmunk}(s_1,s_{1g}) = \left (1 -  \min \left (\frac{\C(s_1,\nmunk)}{\L(s_{1g})}, 1 \right ) \right
) \mychange{* 100}
\label{eq:i_unk}
\end{equation}
where $\min(a,1)$ outputs $1$ when $a \ge 1$, else it outputs the vale of $a$.

So given a set of $w \in W$ words, the patient to be assessed is asked to speak all the
$|W|$ words in $W$; the intelligibility score for the patient is then computed as

\begin{equation} 
\mychange{I_{ld} =} \frac{1}{|W|}\sum_{i=1}^{|W|}I_{sm}(s_i, s_{ig});
\label{eq:final_ld}
\end{equation} 
for Intelligibility based on Sequence Matcher,
\begin{equation} 
\mychange{I_{sm} =} \frac{1}{|W|}\sum_{i=1}^{|W|}I_{ld}(s_i, s_{ig}) ;
\label{eq:final_sm}
\end{equation} 
for Intelligibility based on Levenshtein Distance and
\begin{equation} 
\mychange{I_{\nmunk} =} \frac{1}{|W|}\sum_{i=1}^{|W|}I_{\nmunk}(s_i, s_{ig}) 
\label{eq:final_unk}
\end{equation} 
for Intelligibility based on occurrence of \unk. Note that all these metrics
for intelligibility assessment work, as will be shown in our experiments,  independent of the speaker.

Clearly the larger the number of words ($|W|$) that the patient needs to speak
the more reliable is the intelligibility score, however a
large set of words makes the assessment system poor from the user experience
perspective especially in terms of time required for the patient to speak and
additionally the effort to speak a large number of words by the dysarthric
patient. 
Next we propose a cost minimization approach to select an optimal
number of words (much smaller than $|W|$) without effecting the intelligibility assessment.

\section{Selection of Optimal set of words}
\label{sec:word_selection}

As seen earlier the larger the number of words that the patient is asked to
speak the better the integrity of the intelligibility score. However, from the
user experience and time perspective a smaller number of words is preferable. 
We propose a cost minimization approach to select an optimal number of words from the
original set of $\N$ words that can reliably estimate the subject's intelligibility
instead of having the patient to speak all the $\N$ words. This not only makes the
system less intrusive, but also causes less discomfort to the patient undergoing 
the assessment. 

In order to obtain a subset of the original $\N$ words to be used for the assessment
we identify all possible subsets of $r$ words from the original set of words 
$\SS = \{w_1, w_2, ..., w_{\N}\}$, where $w_{i}$ is the individual word spoken in the utterance 
and $\N$ is the total number of words. The total number of subsets that can be formed from $\SS$ is,
\[
\T = \sum_{r=1}^{\N} {}^{\N}C_r
\]
where 
\[{}^{\N}C_r = \frac{\N!}{\left(\N-r\right)!r!}\] 
is the number of ways of choosing $r$ unordered outcomes from $\N$ possibilities.
As one can see there are $\T$ subsets that can be constructed in all from $\N$
words.

For each subset $\SS_{k}$ namely, $\SS_k \subset W$, 
for $k = \{1,2, ... \T\}$, 
let $|\SS_k|$ denote the number of words in the set $\SS_k$. 
For each subset $W_k$ the intelligibility score is computed as 
\[
I^{\SS_k}_{avg} = \frac{1}{|\SS_k|} \sum_{w \in \SS_k} I_{{w}}
\]
where $I_{{w}}$ is the intelligibility score (for example (\ref{eq:final_ld}))  of the word $w \in \SS_k$ and $I^{\SS_k}_{avg}$ 
denotes the average intelligibility score over all the $|\SS_k|$ words in that subset
$\SS_k$. Note that  $I^{\SS_k}_{avg}$ is computed for a speaker.

Now, we define a correlation 
score of each subset, namely, $I^{\SS_k}_{avg}$ of
a speaker and the corresponding perceptual intelligibility (human annotated
reference) score, $I_{p}$ of
the same speaker \mychange{by computing the Pearson correlation {(\PC)}} as 
\[
\mbox{{\tt{cor}}} \left(\SS_{k} \right) = \mbox{{\mychange{\PC}}}\left(I^{\SS_k}_{avg}, I_{p}\right)
\]
We compute $\mbox{{\tt{cor}}} \left(\SS_{k} \right)$ for all speakers in the
dataset. For sake of simplicity, let the representation $\mbox{{\tt{cor}}} \left(\SS_{k} \right)$
denote the correlation for all the speakers in the database.

To find the optimal number of words, we define the following cost function 
\begin{equation}
\mbox{{\tt{cost}}}\left(\SS_{k}\right ) = 
\alpha_1 |\SS_k| - 
\alpha_2  \mbox{{\tt{cor}}} \left(\SS_{k} \right) 
- \alpha_3 \E \left(\SS_{k}\right)
\label{eq:cost}
\end{equation}
where 
$\E$ is defined as the effort or difficulty in 
pronouncing the words in $\SS_k$. We show, using visible speech \cite{VSbook} analysis 
how to compute $\E$ in
Section \ref{sec:vs_basis}.
Note that  we chose $\alpha_1 = \alpha_2 = \alpha_3 =1$ in all our experiments. 
The subset $\SS^*_k$ that minimizes (\ref{eq:cost}) is the optimal set, namely,
\begin{equation}
\SS_k^* = \min_{k=1,2, \cdots, \T} (\mbox{{\tt{cost}}}\left(\SS_{k}\right )).
\label{eq:mincost}
\end{equation}
As will be seen in the experimental results section, the choice of different
intelligibility metrics results in a different optimal subset $\SS_k$.  

Notice that the cost function (\ref{eq:cost}) has three components, the first
component encourages selection of as few words as possible (from a list of
words or vocabulary), the second
component dependents on the availability of an annotated database and 
encourages selection of those words which show high correlation between
the chosen intelligibility metric and real intelligibility (perceptual) score and the third
component is dependent on the pronunciation of the word and encourages choice
of words (in the vocabulary) that are hard to articulate. The vocabulary is
constrained by the number of (spoken) words that are present in the database and
have been annotated (given a perceptual intelligibility score).

\subsection{Computing Word Pronunciation Difficulty using
Visible Speech}
\label{sec:vs_basis}

Visible Speech (VS) is a system of phonetic symbols 
\cite{VSbook} to represent the 
position of the speech organs in articulating sounds. 
It is composed of symbols that visually capture 
the position and movement of different parts of the speech production 
system. A set of $8$ organs, namely,  
	(a) Larynx,
	(b) Pharynx,
	(c) Soft Palate,
	(d) The action of the soft palate in closing the Nasal Passage,
	(e) Back, (f) Front, (g) Point  of the Tongue,
and 
	(h) Lips
are involved in the speech production system.

The production of speech is a complicated process that involves precise 
control over these $8$ organs. In order to utter a word, phrase or a sentence
of intelligible speech, the rapid switching of positions of different organs 
is of extreme importance. 
The fundamental 
principle of VS is that all the relations between different 
sounds are presented in a symbolic manner. Each symbol captures the position of
the articulators in producing a sound.
The sounds of 
same nature produced at different parts of the mouth are represented by a 
single symbol and the orientation of the symbol depicts the position of the 
organs involved in production of the sound. Based on this hypothesis, he 
proposed a set of Physiological Symbols corresponding to the English elements 
of Speech. Tables \ref{tab:chart} represents the Visible Speech symbols 
corresponding to the consonants, vowels, glides and diphthongs in the English 
language.
\begin{table}
	\caption{Chart of English Sounds and corresponding Visible Speech symbol
\cite{VSbook}}
	\centering
\scalebox{0.7}{
	\begin{tabular}{llllllll}
		&&\multicolumn{5}{c}{Visible Speech Consonants.}& \\
		\vis\ls &p in pea & \vis\ps & t in tea & \vis\bs & k in key & \vis\pp & r in train \\ 
		\vis\lsv &b in bay & \vis\psv & d in day & \vis\bsv & g in gay & \vis\ppv & r in rain \\ 
		\vis\lnv &m in some & \vis\pnv & n in son & \vis\bnv & ng in sung & \vis\tp & h in hue \\ 
		\vis\ld &f in fine & \vis\pb & th in thigh & \vis\pd & l in cloud & \vis\tpv & y in you \\
		\vis\ldv &v in vie & \vis\pbv & th in thy & \vis\pdv & l in loud & \vis\fp & h in hop \\
		\vis\lm &wh in whey & \vis\pmv & s in hiss & \vis\tm & sh in rush & \vis\lmv & w in way \\ 
		\vis\pm &s in his & \vis\tmv & ge in rouge &  &  &  & \\ 
		&&\multicolumn{5}{c}{Visible Speech Vowels.}& \\
		\vis\vhfp &ee in eel & \vis\vhfw & i in ill & \vis\vlfp & e in shell & \vis\vlfw & a in shall \\ 
		\vis\vhbpr &oo in pool & \vis\vhbwr & u in pull & \vis\vlbpr & a in all & \vis\vlbwr & o in doll \\ 
		\vis\vmbw &a in father & \vis\vmmw & a in ask & \vis\vlmw & u in curl & \vis\vmbp & u in dull \\ 
		&&\multicolumn{5}{c}{Visible Speech Glides.}& \\
		\vis\glipback &w in now & \vis\gpoint & r in sir & \vis\gtop & y in may & \vis\gvoice & a in near \\ 
		&&\multicolumn{5}{c}{Visible Speech Diphthongs.}& \\
		\vis\vmbw\gtop &i in mine & \vis\vmfp\gtop & a in mane & \vis\vmbw\glipback & ow in now & \vis\vmbpr\glipback & ow in know \\ 
		 \vis\vlbpr\gtop & oy in boy & & & & & & \\
	\end{tabular}
}
		\label{tab:chart}	
\end{table}
\begin{table}
	\caption{The $10$ Radical Symbols (VS basis) from which all Letters are formed.}
	\centering
\scalebox{0.8}{
	\begin{tabular}{|c|l|}
		\hline
		{VS basis} & {Description} \\ \hline
		\vis\vlmp & The Throat sounding [Voice] \\ 
		\vis\vlmpr & The Throat sounding and lips 'rounded'\\ 
		$\mydot$ & Vowel Definer \\ 
		$\myhat$ & Wide Vowel Definer \\ 
		\vis\fP & The Throat contracted [Whisper] \\ 
		\vis\bp & Part of the Mouth contracted \\ 
		\vis\bd & Part of the Mouth divided \\ 
		$\mymixer$ & Mixer \\ 
		$|$ & Shutter \\ 
		\vis\mnasal & The Nasal valve open [Soft Palate] \\ \hline
	\end{tabular}
}
	\label{tab:10symbols}	
\end{table}
Further, a set of $24$ radical symbols that represent all possible phonetic 
sounds was proposed in \cite{VSbook}. However, of the  $24$ radical symbols, 
only $10$ radical symbols are sufficient to represent all possible vowels and 
consonant letters in the English language. We can look at these $10$ radical
symbols as being a {\em basis} to represent any VS symbol in Table \ref{tab:chart}. 
The $10$ dimensional VS basis (or radical symbols) are 
depicted in Table \ref{tab:10symbols} (some symbols may not accurately depict the radical symbols as proposed by Bell \cite{VSbook}). 
Note that every Visible Speech symbol in Table \ref{tab:chart} 
can be represented in the form of a $10$ dimensional vector, namely, 
[{\vis\vlmp},  {\vis\vlmpr},  $\mydot$,  $\myhat$, {\vis\fP},  {\vis\bp},  {\vis\bd},  $\mymixer$,  $|$,  {\vis\mnasal}].
For example the VS symbol corresponding to k is {\vis\bs} and is formed by
{\vis\bp} and $|$ hence it can be represented as [0, 0, 0, 1, 0, 0, 0, 0, 1, 0]
using the VS basis.  A few VS symbols represented
using the VS basis is shown for ease of understanding in Table
\ref{tab:example}. 

\begin{table}
\caption{Example of VS symbols represented using VS basis.}
\label{tab:example}
\centering
\scalebox{0.8}{
\begin{tabular}{c|rc} \hline
VS Symbol & \multicolumn{2}{c}{VS basis [{\vis\vlmp},  {\vis\vlmpr},  $\mydot$,
$\myhat$, {\vis\fP},  {\vis\bp},  {\vis\bd},  $\mymixer$,  $|$,  {\vis\mnasal}]} \\ \hline
 {\vis\bs}  &  {\vis\bp} and $|$ = &[0, 0, 0, 0, 0, 1, 0, 0, 1, 0] \\
 {\vis\bsv}  & {\vis\bp}, $|$ and {\vis\vlmp} = &[1, 0, 0, 0, 0, 1, 0, 0, 1, 0]\\
 {\vis\pnv}  & {\vis\bp}, {\vis\mnasal} and {\vis\vlmp}= &[1, 0, 0, 0, 0, 1, 0, 0, 0, 1]
\\
 {\vis\pbv} & {\vis\bpv}, $\mymixer$ and {\vis\vlmp}= &[1, 0, 0, 0, 0, 1, 0, 1, 1, 0]
\\
 {\vis\vhbpr} & {\vis\vlmpr} and $\mydot$= &[0, 1, 1, 0, 0, 0, 0, 0, 0, 0]
\\
 {\vis\vhbwr} & {\vis\vlmpr} and $\myhat$= &[0, 1, 0, 1, 0, 0, 0, 0, 0, 0]\\
\hline
\end{tabular}
}
\end{table}
\begin{figure}
	\centering
	\includegraphics[width=0.8\textwidth]{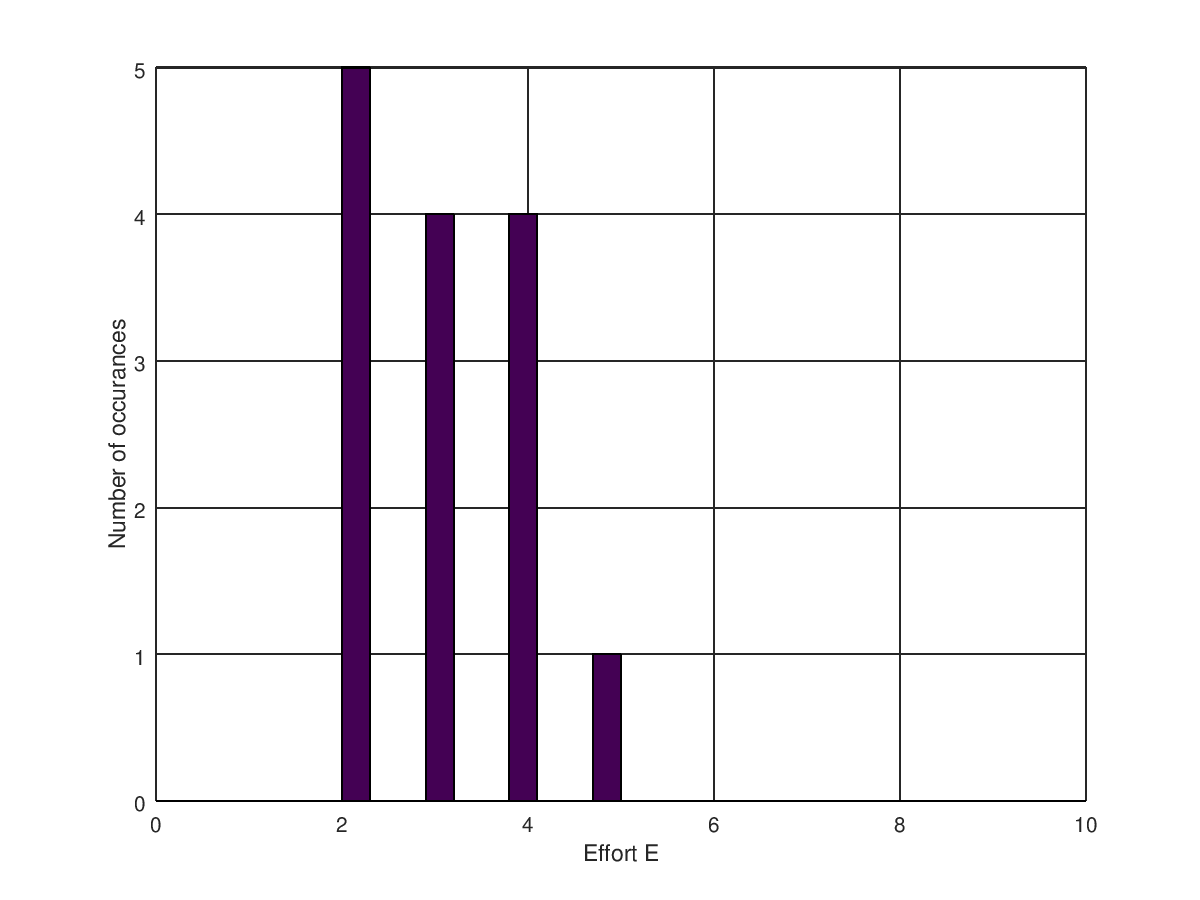}
	\caption{Histogram of the effort involved in production of the word {\naturalization}.}
	\label{fig:naturalization}
\end{figure}

The entire process of decomposition for the English word {\naturalization} is
shown in  Table \ref{tab:naturalization}. Note that  
{\vis\pnv\vlfw\tmv\vhbwr\ppv\vmbp\pdv\vmbp\gtop\pm\vlfp\tm\vmbp\pnv} in the VS symbolic form of
the word {\naturalization} and each VS symbol can be represented as a $10$
dimensional vector using the VS basis.
\begin{table}
	\caption{VS basis representations of the word {\naturalization}.
Note that $\E=43$ is the difficulty level of the word. }
	\label{tab:naturalization}
	\centering
\scalebox{0.7}{
	\begin{tabular}{|c|llllllllll|c|}
		\hline
		 & {\vis\vlmp} & {\vis\vlmpr} & $\mydot$ & $\myhat$& {\vis\fP} & {\vis\bp} & 
{\vis\bd} & $\mymixer$ & $|$ & {\vis\mnasal} & {$E$} \\ \hline
		Rest & [0 & 0 & 0 & 0 & 0 & 0 & 0 & 0 & 0 & 0] & -- \\ 
		{\vis\pnv} & [1 & 0 & 0 & 0 & 0 & 1 & 0 & 0 & 0 & 1] & \textbf{3} \\ 
		{\vis\vlfw} & [1 & 0 & 0 & 1 & 0 & 0 & 0 & 0 & 0 & 0] & \textbf{3} \\
		{\vis\tmv} & [1 & 0 & 0 & 0 & 0 & 1 & 0 & 1 & 0 & 0] & \textbf{3} \\ 
		{\vis\vhbwr} & [0 & 1 & 1 & 0 & 0 & 0 & 0 & 0 & 0 & 0] & \textbf{5} \\ 
		{\vis\ppv} & [1 & 0 & 0 & 0 & 0 & 1 & 0 & 0 & 0 & 0] & \textbf{4} \\ 
		{\vis\vmbp} & [1 & 0 & 1 & 0 & 0 & 0 & 0 & 0 & 0 & 0] & \textbf{2} \\ 
		{\vis\pdv} & [1 & 0 & 0 & 0 & 0 & 0 & 1 & 0 & 0 & 0] & \textbf{2} \\ 
		{\vis\vmbp} & [1 & 0 & 1 & 0 & 0 & 0 & 0 & 0 & 0 & 0] & \textbf{2} \\ 
		{\vis\gtop} & [1 & 0 & 0 & 0 & 0 & 1 & 0 & 0 & 0 & 0] & \textbf{2} \\ 
		{\vis\pm} & [0 & 0 & 0 & 0 & 0 & 1 & 0 & 1 & 0 & 0] & \textbf{2} \\ 
		{\vis\vlfp} & [1 & 0 & 0 & 1 & 0 & 0 & 0 & 0 & 0 & 0] & \textbf{4} \\ 
		{\vis\tm} & [0 & 0 & 0 & 0 & 0 & 1 & 0 & 1 & 0 & 0] & \textbf{4} \\ 
		{\vis\vmbp} & [1 & 0 & 1 & 0 & 0 & 0 & 0 & 0 & 0 & 0] & \textbf{4} \\ 
		{\vis\pnv} & [1 & 0 & 0 & 0 & 0 & 1 & 0 & 0 & 0 & 1] & \textbf{3} \\ \hline
 & & &&&&&&&&&$\E=43$ \\ \hline
	\end{tabular}
}
\end{table}
We now compute the effort ($E$) in terms of  moving the organs of speech production system when
transitioning from one sound represented in the VS basis as $\vec{R_i}$ to the next sound represented
in the VS basis as ($\vec{R_k}$) as 
\[
E = \#({\vec{R_{i}} \oplus \vec{R_{k}}})
\]
where, $\oplus$ represents the XOR operation and $\#(\vec{R})$ represents the number of
$1$'s in the vector $\vec{R}$. The last column in Table \ref{tab:naturalization}
shows the effort, $E$, in traversing from one sound to another. For example,
$E=3$ for traversing from sound {\vis\pnv} represented by  
$R_{\mbox{\vis\pnv}}$ 
=[ 1, 0, 0,  0, 0, 1, 0, 0, 0, 1]
to  the sound {\vis\vlfw} represented by $R_{\mbox{\vis\vlfw}}$ 
= [1, 0, 0, 1, 0, 0, 0, 0, 0, 0] (see Table \ref{tab:naturalization}).
Note that $\vec{R_{\mbox{\vis\pnv}}} \oplus \vec{R_{\mbox{\vis\vlfw}}}$ 
$= [ 0,  0,  0,  1,  0,  1,  0,  0,  0,  1$] and $\#([ 0,  0,  0,  1,  0,  1,  0,
0,  0,  1]) = 3$. The effort  for pronouncing the word {\naturalization}
is the sum of effort taken to traverse from one sound to another while
speaking the word. As seen in Table \ref{tab:naturalization} the effort to
speak naturalization is $\E=43$.

Figure \ref{fig:naturalization} shows the histogram plot for the effort involved in producing 
the word {\naturalization}. Observe that there are $5$ instances when an
effort or $E=2$ was required to move from one sound to another. 
Since, the ability to precisely control the different organs involved in speech production 
is diminished in dysarthric patients, words that require higher value of effort are 
usually not articulated properly by these patients. Subsequently words with
higher $\E$ are more difficult to articulate because they need speech producing
organs to transition more than a simple word. 

The effort required to speak a word or equivalently the difficulty level of a
word depends on the amount of changes the articulators have to make to
articulate that word. We use this in building a cost function (\ref{eq:cost})
to identify an
optimal set of words that can be used to assess the intelligibility of spoken
speech.

\section{Experimental Results}
\label{sec:experiments}

We conducted a series of experiments with an objective to determine how close
the proposed intelligibility assessment techniques were to the SLP relatable 
perceptual intelligibility score. Additionally the aim was to assess the
intelligibility score by asking the dysarthric patient speak a very limited
number of {\em carefully chosen} words. We first detail the database that was
used for experimentally validating the proposed speech intelligibility
techniques.

\subsection{Database} 
\label{section:database}

The Dysarthric Speech Database for Universal Access Research (UA Speech database) 
\cite{UASpeech} consists of audiovisual recordings produced by $15$ dysarthric 
speakers with cerebral palsy and $13$ healthy subjects. The subjects were 
required to read a set of $765$ isolated words \mychange{($449$ unique)} shown on a computer monitor. 
The set of utterances, from each participant, included
\begin{itemize}
	\item \Digits\ (D) ($10$ words, $3$ repetitions).  Example {\tt
one}, {\tt two} , $\cdots$
	\item \Letters\ (L) ($26$ words, $3$ repetitions).
The $26$ letters of the International Radio Alphabet. Example {\tt alpha},
{\tt bravo}, $\cdots$
	\item \CC\ (CC) ($19$ words, $3$ repetitions). 
A set of common word processing commands. Example {\tt command}, {\tt line}, 
{\tt enter}, $\cdots$ 
	\item \CW\ (CW) ($100$ words, $3$ repetitions). 
The most common $100$ words taken from the Brown corpus. Example {\tt {yes}}, {\tt {no}}, $\cdots$
	\item \UW\  (UW) ($300$ words, $1$ repetition). 
Words selected from Project Gutenberg novels such that the occurrence of 
infrequent biphones is maximized. Example {\tt {butterflies}}, {\tt {convulsion}}, $\cdots$
\end{itemize}
\mychange{As can be observed there should be $455$ ($10$ \Digits, $26$ \Letters, $19$
\CC, $100$ \CW, and $300$ \UW) unique words. However $6$
words ({\tt choking}, {\tt equilibrium}, {\tt moustache}, {\tt powwow}, {\tt vouchsafe}, and {\tt watch}) in \UW\
repeat reducing the number of unique words to $449$.}

\begin{table}
	\caption{Distribution of speaking task into blocks B1, B2 and B3. The number of utterances is shown in $\{\}$ and the repetition number is shown $()$
\cite{9053339}.}
	\begin{center}
	\scalebox{0.8} 	{
		\begin{tabular}{l|c||c|c|c}\hline
			&Full $\{765\}$& B1 $\{255\}$& B2 $\{255\}$& B3 $\{255\}$\\ \hline\hline
		\CC\ &CC $\{57\}$ & CC(1) $\{19\}$ & CC(2) $\{19\}$ & CC(3)  $\{19\}$\\
		\Letters\	&L $\{78\}$ & L(1) $\{26\}$ & L(2) $\{26\}$& L(3) $\{26\}$\\
		\Digits\	&D $\{30\}$ & D(1) $\{10\}$ & D(2) $\{10\}$& D(3) $\{10\}$\\
		\CW	&CW $\{300\}$ & CW(1) $\{100\}$& CW(2) $\{100\}$& CW(3) $\{100\}$\\
	\UW		&UW $\{300\}$ & UW1 $\{100\}$ & UW2 $\{100\}$ & UW3 $\{100\}$\\ \hline
		\end{tabular}
}
	\end{center}
	\label{tab:UAblocks}
\end{table}

All utterances, as mentioned in \cite{UASpeech}, were recorded using an eight-microphone array at a 
sampling frequency of $48$ kHz. For ease of subjects, the words were divided into three equally 
sized blocks (B1, B2, B3) of $255$ words each and the participants were given a break between the 
blocks. The distribution of \mychange{utterances} in the three blocks is shown in Table \ref{tab:UAblocks}
for each speaker \cite{9053339}. 

A set of $5$ native American English were asked to provide orthographic transcriptions for utterances 
spoken by dysarthric patients in order to assign a perceptual intelligibility rating. For each 
listener's transcription, the percentage of correctly identified transcription was calculated and 
averaged to obtain the individual speaker's intelligibility rating based on perception. 
These intelligibility scores lie in the range of $2\%$ to $95\%$. Further, based  on  these 
scores,  each speaker was classified into one of four categories: \VeryLow\ $(0\%-25\%)$, 
\Low\ $(26\%-50\%)$, \Medium\ $(51\%-75\%)$ and \High\ $(76\%-100\%)$
intelligibility.

\subsection{Results}
As an attempt to concurrently seize the effects of the proposed techniques, 
we evaluate the performance of the methods, namely, 
(i) $I_{os}$ (\ref{eq:i_os}),
(ii) \DS\ + $I_{sm}$ (\ref{eq:i_sm}), 
(iii) \DS\ + $I_{ld}$ (\ref{eq:i_ld}) and
(iv) \DS\ + $I_{\nmunk}$ (\ref{eq:i_unk}) 
on different subsets of utterances from the UASpeech dataset. 
As mentioned earlier $I_{os}$ is negatively correlated to perceptual intelligibility scores
while the other intelligibility scores are positively correlated. 
The subsets used for our analysis, as seen in  Table \ref{tab:UAblocks}, are
(a)  \CC\ (CC), (b) International Radio Letters (\Letters), (c) \Digits, (d) \CW\  (CW) and 
(d) \UW\ UW1, UW2 and UW3 and  across three different blocks,
namely, B1, B2 and B3 as shown in Table \ref{table:detailedres}.
We also compute the combined average score across all $765$ words using our intelligibility 
metrics $I_{os}$, \DS\ + $I_{sm}$, \DS\ + $I_{ld}$, and \DS\ + $I_{\nmunk}$ compared to the perceptual intelligibility (considered
as the ground truth) rating using \PC. Since all but \UW\ are present in each 
of the three blocks B1, B2 and B3, the \PC\ values averaged across the three blocks are of the same range. 
The low value of \PC\ for the subset of \Digits\ is attributed to the comparative simplicity in production 
of the words and hence less effort to pronounce them as described by the visible speech analysis. As
we will see later, none of the words in \Digits\ qualify into the optimal set of
words that are sufficient to compute the intelligibility score of dysarthric
speech.
In general the \UW\  are difficult to pronounce and hence the higher value of \PC\ signifies that the UW subset 
seems to captures the nuances required for reliable intelligibility estimation.
It can also be seen that intelligibility score based on Sequence Matching
and Levenshtein
distance performs better compared to the other intelligibility measuring
metrics.

\begin{table}
	\caption{Pearson Correlation of the proposed techniques for assessment of intelligibility.}
	\centering
	\resizebox{\textwidth}{!}{
		\begin{tabular}{|c|c|c|c|c|c|c|c||c|c|c||c|} \hline
			& CC & \Digits\ & \Letters & CW & UW1 & UW2 & UW3 & B1 & B2 & B3 &	All Words\\ \hline			
			Utt/speaker & 57 & 30 & 78 & 300 & 100 & 100 & 100 & 255 & 255 & 255 & 765\\ \hline
			$I_{os}$ (\ref{eq:i_os})& -0.80 & -0.74 & -0.83 & -0.81 & -0.80 & -0.73 & -0.74 & -0.84 & -0.80 & -0.73 & -0.81\\ \hline
		\DS\ + 	$I_{sm}$ (\ref{eq:i_sm})& 0.85 & 0.84 &	0.93 & 0.94 & 0.94 & 0.92 &	0.92 & 0.94 & 0.93 & 0.94 & {\bf 0.94}\\ \hline
		\DS\ + 	$I_{ld}$ (\ref{eq:i_ld})& 0.88 & 0.84 & 0.94 & 0.93 & 0.91 & 0.90 & 0.90 & 0.93 & 0.91 & 0.93 & 0.93 \\ \hline
		\DS\ + 	$I_{\nmunk}$(\ref{eq:i_unk}) & 0.87 & 0.83 & 0.88 & 0.82 & 0.91 & 0.90 & 0.90 & 0.89 & 0.88 & 0.87 & 0.88\\ \hline
		\end{tabular}}
		\label{table:detailedres}
	\end{table}

As we argued earlier that it is physically taxing on a patient to speak a large
number of utterances, for example, the best performance of intelligibility assessment
would require the patient to speak all the $765$ \mychange{utterances} (last column of Table
\ref{table:detailedres}). We applied the cost minimization approach, described in
Section \ref{sec:word_selection}, to identify an optimal (much smaller than
\mychange{$449$}) set of words, which would not adversely affect the computation of intelligibility score of the
speaker. 

Given a set of $\N$ words, the proposed algorithm requires to look in for $(2^{\N} - 1)$ possible 
combinations to find the optimal subset $\SS_k$ containing  $\r$ words. 
In the UASpeech dataset a set of $\N = \mychange{449}$ \mychange{unique} words are provided which leads to $\T
 \approx 10^{\mychange{135}}$ possible subsets. Note that operating on such high number of
possible solutions is infeasible. We used the following two criteria to reduce
the number of words from which we expect our word selection to happen

\begin{enumerate}
	\item All words with 
less than $\mychange{5}$ syllables are discarded.
	\item For each isolated word, the distance traversed in the 2D vowel space 
(formed by formants $F_1$ and $F_2$) is determined using the standard values of formants for 
vowels shown in Table \ref{tab:formantvalues}
\cite{formantvalues2, formantvalues}.
The phonetic transcription of each word is obtained and the Euclidean distance between subsequent vowels 
present in the word are summed up to obtain the distance that is required to be traversed in the vowel 
space in order to satisfactorily pronounce the word. 

For example, the word {\naturalization} has the phonetic transcription (using
the ARPABET symbol set consists of $39$ phonemes \cite{arpabet_phone})
{['N', 'AE', 'CH', 'ER', 'AH', 'L', 'IH', 'Z', 'EY', 'SH', 'AH', 'N']}, considering only 
the vowels, 
we get {['AE', 'ER', 'AH', 'IH', 'EY', 'AH']}.
Assuming an initial rest start, namely, $F_1 = 0$ and $F_2 = 0$, for the word 
the distance traversed in the
$F_1-F_2$ space is calculated as the Euclidean distance between the vowels,
namely, 
'AE' (588, 1952) $\rightarrow$ 
'ER' (474, 1379) $\rightarrow$ 
'AH' (623, 1200) $\rightarrow$ 
'IH' (427, 2034) $\rightarrow$ 
'EY' (580, 1799) $\rightarrow$
'AH' (623, 1200) is {$4593.44$}. All words that traversed less than $2400$ in
this $F_1-F_2$ space were discarded.

\end{enumerate}   
	
	\begin{table}
		\caption{Formant frequencies of F1 and F2 for American English vowels from \cite{formantvalues}. }
		\centering {
			\scalebox{0.8}{
		\begin{tabular}{|l|l|l|l|l|l|l|l|l|l|l|l|} \hline
{{Vowel}} & {{E{Y}}} & {AO} & {ER} & {AH} & {UW} & {AE} & {IY} & {IH} & {UH} & {AA} & {E} \\ \hline 
{{F1}}    & {580}  & 652  & 474  & 623  & 378  & 588  & 342  & 427         & 469         & 768         & 476 \\ \hline 
{{F2}}    & {1799}        & 997         & 1379        & 1200        & 997         & 1952        & 2322        & 2034        & 1122        & 1333        & 2089 \\ \hline 
		\end{tabular}
	}
}
\label{tab:formantvalues}
	\end{table}

\mychange{
The aforementioned constraints  reduces the space of 
words from \mychange{$449$} to \mychange{$14$}. 
The set of $14$ words, all from \UW\ set are shown in Table
\ref{tab:14words}. 
We applied the proposed cost minimization approach on
this set of \mychange{$14$} words only. Note that we now have to search for an
optimal set of words from $\T = 16384$ possible subsets (see
(\ref{eq:mincost})). 
The performance, in terms of \PC, of the proposed techniques for
assessment of intelligibility on these set of $14$ words is also shown in Table 
\ref{tab:14words}. Note that this is our reference performance of the four
intelligibility techniques. We now apply our cost minimization approach to find
the optimal set of words from this set of $14$ words (Table
\ref{tab:14words}).}
\begin{table}
\caption{\mychange{Performance on the set of $14$ words identified from the set
of $449$ words, which had more than $5$ syllables and traversed $> 2400$ in the
$F_1$-$F_2$ space.}}
\label{tab:14words}
\centering
\scalebox{0.8}{
\begin{tabular}{|l|c|c|p{7.5cm}|} \hline
Method & $\r$ & \PC & 
\multirow{5}{*}{\vbox{{
\NATURALIZATION, \AUTOBIOGRAPHY,
\EXACTITUDE, \IRRESOLUTE, \INALIENABLE, \LEGISLATURE, \OVERSHADOWED,
\PSYCHOLOGICAL, \DISSATISFACTION, \AGRICULTURAL, \APOTHECARY,
\AUTHORITATIVE, \EXAGGERATE, \INEXHAUSTIBLE  
}}}\\ \cline{1-3}
$I_{os}$ & 14 &  -0.81 & \\ \cline{1-3}
\DS\ + $I_{\nmunk}$ & 14 &  0.91 & \\ \cline{1-3}
\DS\ + $I_{sm}$ & 14 &  0.92  &\\ \cline{1-3}
\DS\ + $I_{ld}$ & 14 &  0.87  &\\  \hline
\end{tabular}
}
\end{table}

\mychange{
In the experiments to follow, we minimize the cost function (\ref{eq:cost}) for
different values of $\alpha$'s to identify an optimal set of words ($\SS_k^*$) 
and show the performance on this set of words for the four different
intelligibility assessment metrics, namely, $I_{os}$, \DS\ + $I_{sm}$, \DS\ + $I_{ld}$, \DS\
+ $I_{\nmunk}$. 
\begin{theorem}
$
\SS_1^* = \min_{k=1,2, \cdots, \T} 
\left \{\alpha_1 |\SS_k| 
- \alpha_3 \E \left(\SS_{k}\right) \right\}$
\end{theorem}
This is the case when
we do not have access to perceptual intelligibility score, or in other words
there is no access to a database of utterances. Clearly,
this scenario is (a) independent of the intelligibility assessment 
technique described in Section \ref{sec:proposed} (see Table \ref{tab:optimal_101}) and (b) allows for building a
speech intelligibility assessment system for any language using just the
dictionary of words without the need for an annotated database. }

\mychange{
The optimal set of words for $\alpha_1=1, \alpha_2=0, \alpha_3=1$ is shown in
Table \ref{tab:optimal_101}. As can be seen there is not a significant change
in the performance of the intelligibility assessment techniques when a set 
of $8$
words were selected from $14$ words using the cost minimization approach with
$\alpha_1=1, \alpha_2=0, \alpha_3=1$ in (\ref{eq:cost}). This suggests that
even in the absence of a database we should be in a position to select a 
set of words that can be used for intelligibility assessment.
} 
\begin{table}
\caption{\mychange{Performance on selected optimal set of words for $\alpha_1=1,
\alpha_2=0, \alpha_3=1$.}}
\label{tab:optimal_101}
\centering
\scalebox{0.8}{
\begin{tabular}{|l|c|c|p{7.5cm}|} \hline
Method & $|\SS_1^*|$ & \PC & Selected Words ($\SS_1^*$)\\\hline
$I_{os}$&  8 & -0.82 &\multirow{4}{*}{\vbox{{\naturalization, \authoritative, \autobiography,
\psychological, \dissatisfaction, \agricultural, \apothecary, \inexhaustible}}}\\ \cline{1-3}
\DS\ + $I_{\nmunk}$ & 8 &  0.93 & \\ \cline{1-3}
\DS\ + $I_{sm}$ & 8 &  0.91  &\\ \cline{1-3}
\DS\ + $I_{ld}$ & 8 &  0.85  &\\  \hline
\end{tabular}
}
\end{table}

\mychange{
\begin{theorem}
$
\SS_2^* = \min_{k=1,2, \cdots, \T} 
\left \{\alpha_1 |\SS_k| 
- \alpha_2  \mbox{{\tt{cor}}} \left(\SS_{k} \right)
\right\}$
\end{theorem}
This condition $\alpha_1=1, \alpha_2=1,$ and $\alpha_3=0$ does not make use of
the component that captures the articulatory effort (based on visual speech) required to 
utter a word. 
The optimal set of words for $\alpha_1=1, \alpha_2=1, \alpha_3=0$ is shown in
Table \ref{tab:optimal_110}. 
Clearly the introduction of effort required to utter a word
(computed using visual speech) is a very important factor in 
the construction of
the cost function. 
As can  be seen, the absence of visual speech based effort to utter a word
in the cost function (\ref{eq:cost}) results in the selection of only one word. 
The selected optimal word happens to be the one with the largest \PC\
depending on the technique used for assessment.
}

\begin{table}
\caption{\mychange{Performance on selected optimal set of words for $\alpha_1=1,
\alpha_2=1, \alpha_3=0$.}}
\label{tab:optimal_110}
\centering
\scalebox{0.8}{
\begin{tabular}{|l|c|c|p{7.5cm}|} \hline
Method & $|\SS_2^*|$ & \PC & Selected Words ($\SS_2^*$)\\\hline
$I_{os}$&  1 & -0.79 &\dissatisfaction\\ \hline
\DS\ + $I_{\nmunk}$ & 1 &  0.89 & \authoritative\\ \hline
\DS\ + $I_{sm}$ & 1 &  0.88  &\naturalization\\ \hline
\DS\ + $I_{ld}$ & 1 &  0.85  &\naturalization\\  \hline
\end{tabular}
}
\end{table}

\begin{theorem}
$
\SS_3^* = \min_{k=1,2, \cdots, \T} 
\left \{\alpha_1 |\SS_k| 
- \alpha_2  \mbox{{\tt{cor}}} \left(\SS_{k} \right)
- \alpha_3 \E \left(\SS_{k}\right) 
\right\}$
\end{theorem}

The optimal value of the number of words ($|\SS_3^*|$), and the subset of words
$\SS_3^*$ for the four different intelligibility assessment techniques is shown in  Table \ref{table:r-PC}. 
Also shown is the \PC\ upon performing the intelligibility assessment with the identified optimal set of words.
It can be observed that the optimal number of words selected are between \mychange{$6$ (for
$I_{os}$) and $9$ (for \DS\ + $I_{sm}$)}. 
This is a significant reduction in the number of utterances compared to the $765$ \mychange{utterances} that a dysarthric patient 
has to speak to enable gauge the intelligibility score (\cite{Paja,Martinez,ICASSP2019janbakhshi}). The automatic identification of an optimal number of
words, using a cost minimization approach, for intelligibility assessment is one of the main contributions of this
paper. 
It can be observed that the selected words, in the optimal set, require complex movements and 
precise control over the organs of the speech production system and hence, are suitable for the 
purpose of intelligibility assessment.
	\begin{table}
		\caption{Performance of the proposed techniques 
after selecting an optimal set words (\ref{eq:cost}) \mychange{for
$\alpha_1=\alpha_2=\alpha_3=1$}.}
		\centering
		{
\scalebox{0.8}{
			\begin{tabular}{|l|c|c|p{150px}|}
				\hline
Method & $|\SS_3^*|$ & \PC & Selected Words ($\SS_3^*$)\\\hline
$I_{os}$ & $6$ & $-0.86$ & {\autobiography, \overshadowed, \psychological,
\dissatisfaction, \agricultural, \inexhaustible}  \\\hline 
\DS\ + $I_{\nmunk}$ & $7$ & $0.94$ & {\naturalization, \psychological,
\dissatisfaction, \agricultural, \apothecary, \authoritative, \inexhaustible}\\\hline 
\DS\ + $I_{sm}$ & $9$ & $0.94$ & {\naturalization, \authoritative, \exactitude,
\overshadowed,
 \psychological, \dissatisfaction, \agricultural, \apothecary,
 \inexhaustible,} \\\hline
\DS\ + $I_{ld}$ & $6$ & $0.91$ & {\inexhaustible, \authoritative, \apothecary,
\agricultural,
 \dissatisfaction, \naturalization}\\\hline 
			\end{tabular}} }
			\label{table:r-PC}
		\end{table}

\begin{figure}[ht]
	\centering
	\includegraphics[width=0.9\textwidth]{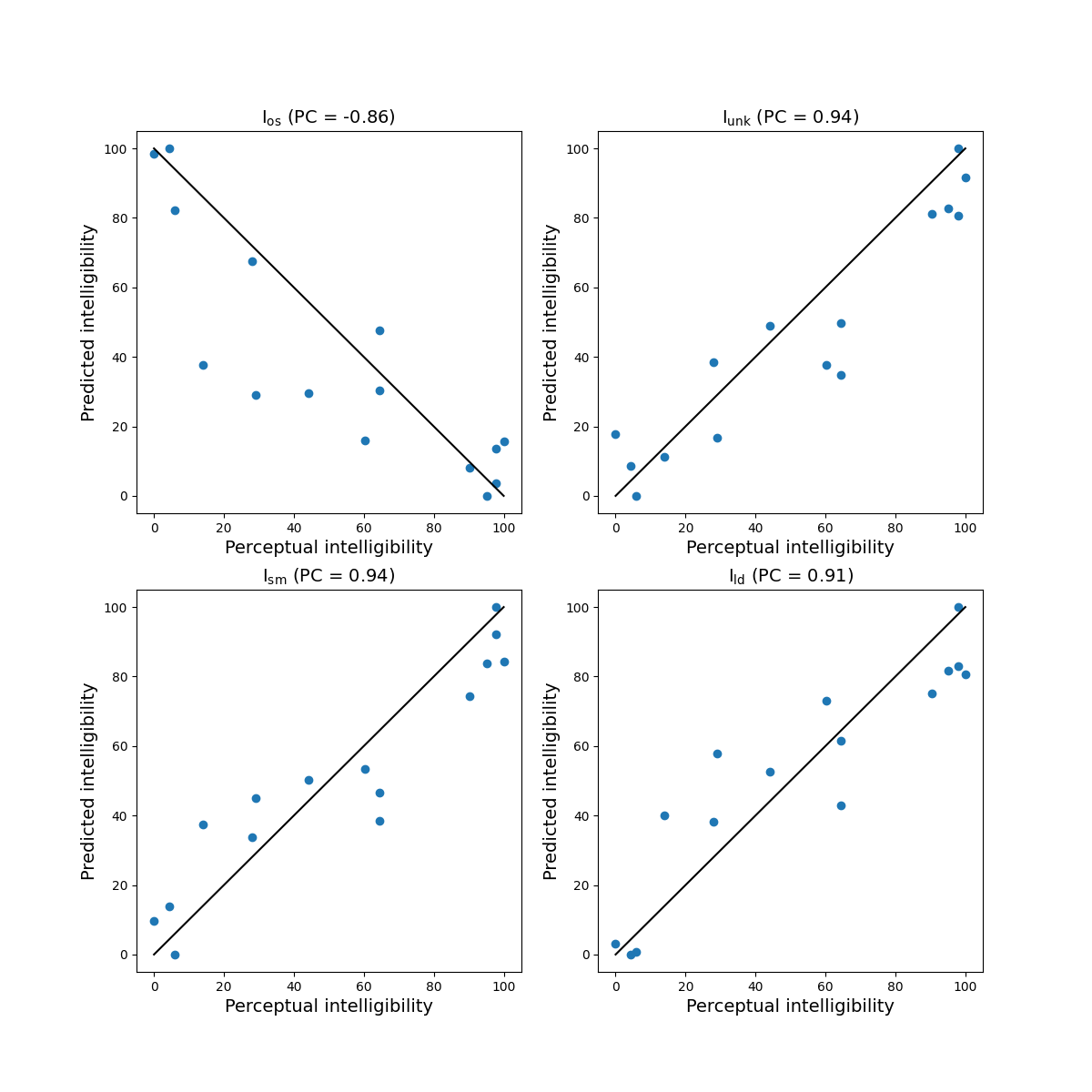}
	\caption{Scatter plot between the perceptual intelligibility ($I_p$) and
$I_{os}$ (top left), \DS\ + $I_{\nmunk}$ (top right), \DS\ + $I_{sm}$ (bottom left) and
\DS\ + $I_{ld}$ (bottom right).}
	\label{fig:plots}
\end{figure}

Figure \ref{fig:plots} shows a scatter plot of the estimated intelligibility rating based on each of the four proposed approaches,
namely, $I_{os}$ (top left), \DS\ + $I_{\nmunk}$ (top right), \DS\ + $I_{sm}$ (bottom left) and
\DS\ + $I_{ld}$ (bottom right) with the perceptual intelligibility rating ($I_p$). 
It can be  inferred that the predicted and perceptual assessment scores follow a linear relation throughout the intelligibility 
range ($0$-$100$). Thus, we establish that even a smaller set of {\em optimally} chosen words is indeed sufficient to reliably 
assess the intelligibility rating of a dysarthric speaker. It can be observed 
 that the \PC\ of
all the proposed techniques (Table \ref{table:r-PC}) based on a set of \mychange{only} $6$
to $9$ spoken words is \mychange{not very far} from the
last column of Table \ref{table:detailedres} which is based on  considering all the $765$ 
\mychange{utterances} spoken by the patient. 

We further evaluate the proposed four different intelligibility assessment systems
with other state-of-the-art techniques in Table \ref{table:comparison}. 
In \cite{Martinez}, the authors obtained a \PC\ of $0.91$ using an approach based on 
a total variability subspace modeled by factor analysis by representing acoustic information 
corresponding to each speech by an i-vector and then trained a support vector regression 
model for intelligibility score estimation. Janbakhshi et. al. \cite{ICASSP2019janbakhshi} 
proposed a reference-based approach called P-STOI by using DTW technique to create utterance 
dependent healthy references and then computing spectral cross-correlation of aligned 
pathological speech to the healthy reference in order to assess intelligibility, yielding a \PC\ of $0.95$. 
A Mahalanobis distance-based discriminant analysis classifier based on a set of acoustic features, 
resulted in a \PC\ of $0.96$ as proposed in \cite{Paja}.
As can be observed, all the \DS\ based proposed approaches of assessing the 
intelligibility of dysarthric speech \mychange{are very close to}
(in terms of correlating with perceptual intelligibility score) other state of
the art approaches proposed in literature in spite of the fact that our
approaches require a significantly smaller number of words to be uttered by the
patient (compare $6-9$ utterances to $765$ utterances). 

\begin{table}
	\caption{Performance of the proposed and state-of-the-art measures on the UA Speech corpus.}
	\centering
	\scalebox{0.85}
	{
		\begin{tabular}{|l|c|} \hline
			Method & \PC\\ \hline \hline
			Martinez et. al.\cite{Martinez} (2013) & $0.91$\\ \hline
			Janbakhshi et. al.\cite{ICASSP2019janbakhshi} (2019) & $0.95$\\ \hline
			Paja et. al.\cite{Paja} (2012) & $0.96$\\ \hline \hline
			$I_{os}$ (Proposed) & \mychange{$-0.86$}\\ \hline
			\DS\ + $I_{\nmunk}$ (Proposed) & \mychange{$0.94$} \\ \hline
			\DS\ + $I_{sm}$ (Proposed) & \mychange{$0.94$}\\ \hline
			\DS\ + $I_{ld}$ (Proposed) & \mychange{$0.91$}\\ \hline
		\end{tabular}}

		\label{table:comparison}
	\end{table}

\mychange{
It can be observed that the word \naturalization\ appears frequently as one of
the words in the optimal set of words selected by minimizing the cost function
(\ref{eq:cost}) for varying values of $\alpha$'s (see Tables \ref{tab:optimal_101},
\ref{tab:optimal_110},  \ref{table:r-PC}). The performance of the
proposed techniques using the single word \naturalization\ in terms of \PC\ is
captured in Table \ref{tab:frequentword}. The main reason for its frequent
occurrence can be attributed the fact that it has one of the highest articulatory effort cost (computed based
on visual speech) among all the $14$ words (see Table \ref{tab:naturalization})
that were used in our experiments to select the optimal set of words.}
\begin{table}
	\caption{Performance of the proposed techniques on the frequently
occurring word \naturalization.}
\label{tab:frequentword}
	\centering
	\scalebox{0.85}
	{
		\begin{tabular}{|l|c|} \hline
			Method & \PC\\ \hline \hline
			$I_{os}$  & \mychange{$-0.55$}\\ \hline
			\DS\ + $I_{\nmunk}$  & \mychange{$0.85$} \\ \hline
			\DS\ + $I_{sm}$  & \mychange{$0.88$}\\ \hline
			\DS\ + $I_{ld}$  & \mychange{$0.93$}\\ \hline
		\end{tabular}
	}
\end{table} 

\section{Conclusion}
\label{sec:conclusions}

Dysarthria is a neuro motor disorder that affects the ability of the person to
have precise control over his organs that help in producing speech. Determining
the intelligibility of speech which is relatable with the perceptual
intelligibility  score that the SLP is familiar with  is important.
In this paper, we introduced four different techniques, all speaker
independent, that can assist in computing the
intelligibility score of a dysarthric speaker. The techniques make use of the
output of an end to end pre-trained \DS\ speech to alphabet \mychange{(S2A)} engine, which in our
opinion is novel.  As mentioned earlier, we exploit the fact that lack of
intelligibility detoriates the performance of the \mychange{S2A} engine
primarily because it has been trained on healthy normal speech. Keeping in mind, the immense
stress and difficulty faced by a dysarthric patient to speak  a large number of
words, we proposed a cost minimization scheme which allows for identification of
an optimal set of words that are sufficient to retain the fidelity of speech
intelligibility assessment. 
This formulation of the cost minimization approach to identify the optimal set
of words, the use of visual speech to determine the pronunciation difficulty of a word,
in our opinion, has never been explored before. 
Infact the cost minimization approach, as formulated in this paper, can be employed to identify an optimal 
set of words suitable for intelligibility assessment, given a set of dictionary words in any other language
\mychange{(see Table \ref{tab:optimal_101})}. 
Experimental results show that the choice of optimal number of words is able to predict the intelligibility score of a dysarthric patient 
and is linearly proportional to the perceptual intelligibility score, thereby making the 
proposed system built on optimal set of words usable by a speech language pathologist.

\bibliography{rev_sc_dysarthria_intelligibility}
\end{document}